\documentclass[final,3p,times,twocolumn]{elsarticle}
\usepackage{graphicx}
\usepackage{amssymb}
\usepackage{amsmath}
\usepackage{hyperref}
\usepackage{gensymb}

\usepackage{amsmath}
\usepackage{makeidx}
 
 \journal{Physica D}

\usepackage[usenames]{color}
\usepackage[normalem]{ulem}
\usepackage{gensymb}

\usepackage{hyperref}
\usepackage[T1]{fontenc} 
\begin{document}

\title
{Giant vortex in a harmonically-trapped rotating  dipolar  $^{164}$Dy condensate}

\author[fac]{Luis E. Young-S.}
\ead{lyoung@unicartagena.edu.co}

 \address[fac]{Grupo de Modelado Computacional, Docente de Planta, Departamento de Física,  Facultad de Ciencias Exactas y Naturales, Universidad de Cartagena, 
130015 Cartagena de Indias, Bolivar, Colombia}

 \author[int]{S. K. Adhikari}
\ead{sk.adhikari@unesp.br}

\address[int]{Instituto de F\'{\i}sica Te\'orica, UNESP - Universidade Estadual Paulista, 01.140-070 S\~ao Paulo, S\~ao Paulo, Brazil}

\begin{abstract}

We demonstrate the  formation of  dynamically stable giant vortices in a harmonically-trapped  strongly dipolar 
$^{164}$Dy Bose-Einstein condensate   under rotation around the polarization direction of dipolar atoms, employing the numerical solution of  an improved  mean-field model including a   Lee-Huang-Yang-type interaction,  meant to stop a collapse at high atom density. These giant vortices   are stationary,  obtainable by imaginary-time propagation using a  
 Gaussian  initial state, while the appropriate  phase of the  giant vortex is imprinted on the initial wave function.  The dynamical 
stability of the giant vortices is established by real-time propagation during  a long  interval of time   after  a  small change of a parameter.

\end{abstract}

\maketitle

%


\section{Introduction}

\label{I}

After the observation of harmonically-trapped Bose-Einstein condensates (BEC) of alkali-metal atoms at ultra-low temperatures in a laboratory \cite{Expt1,Expt2}, there have been many experiments to investigate their superfluid properties. An earmark of superfluidity is the generation of a quantized vortex  in a rotating BEC. 
 According to { Onsager \cite{onsager},  Feynman \cite{feynman}, and Abrikosov \cite{4}  } these vortices have quantized circulation   \cite{2c}:
\begin{equation} \label{cir}
\frac{m}{2\pi\hbar }\oint_{\cal C}  {\bf v}.d{\bf r}= \pm l,
\end{equation}
where  ${\bf v}({\bf r}, t)$ is superfluid  velocity at  point ${\bf r}$ and time $t$, $ l$ is the
quantized angular momentum of an atom   in unit of $\hbar $, $m$ is its mass,
$\cal C$  
is a generic closed path.  
 For $l\ne 0$,  a quantized vortex line appears \cite{2c} due to topological defects inside the closed path $\cal C$.    
 
 Under slow and rapid rotational velocity, a quantized  vortex  \cite{3a,2a,2b,2c} and  a  vortex-lattice structure  \cite{vorlat} were created, respectively, in a rotating BEC  under   controlled conditions and were 
 experimentally investigated.  With the increase of the angular velocity of rotation of a harmonically-trapped BEC, many vortices of unit angular momentum each per atom ($l = 1$)  are formed on an Abrikosov triangular lattice \cite{4}.    An angular momentum state with $l > 1$ in a single-component harmonically-trapped BEC \cite{3a,vorlat,3b,7a,7c} is unstable from an energetic consideration and decays into multiple vortices of unit angular momentum each.
  In a multicomponent  spinor BEC one can also have the formation of square \cite{10,11a,11b,11b2}, stripe and honeycomb \cite{12} vortex lattice, different from the triangular  Abrikosov  lattice \cite{4}. 
   Moreover,  it is possible to create    coreless
vortices \cite{13},  fractional angular momentum vortices \cite{14a,14b}
and phase-separated vortex lattices in multi-component 
  spinor \cite{mu4} and dipolar \cite{11a,11b} BECs.

In spite of continued efforts \cite{exGV1}, it has not been possible to generate a  giant vortex of large 
angular momentum ($l> 1$) in a harmonically trapped BEC.  In addition to theoretical interest, such a vortex could be useful in  quantum information processing \cite{qi1,qi2}. The existence of a  giant vortex in a BEC confined by a quadratic plus a
quartic trap, \cite{gv0,gv1,gv2,gv5,gv6,gv21,gv7}, a bubble trap \cite{gv10}
or a  Gaussian plus a quadratic trap  \cite{gv8}
have also been theoretically predicted. In all these cases   the effective trapping potential develops a repulsion in the  central region, like a Mexican hat potential,
resulting in a hole along the rotation axis which accommodates a giant vortex.  It was also possible to observe  a  dynamically-formed metastable giant vortex of large angular momentum ($60>l>7$) in a rapidly-rotating harmonic trap \cite{gv3} and investigate it numerically \cite{gv4}. { There has  been significant work  on the creation of giant vortices in a regular BEC as  a negative temperature state \cite{neg}.
A giant vortex in a multicomponent nondipolar \cite{mu1,mu2,mu3} and  dipolar \cite{mud} BEC and also in a spin-orbit coupled BEC \cite{mu3} was also studied.}

In view of the above, it is highly desirable to have a giant vortex in a single-component harmonically-trapped BEC, if that is at all possible. To this end we demonstrate the formation of giant vortices of large  angular momentum ($5\ge l \ge 2$)  in a harmonically-trapped  BEC of $^{164}$Dy atoms,  with large dipole moment, 
 under rotation 
around the polarization direction $z$ of the dipolar atoms.
{Previously,  significant work was done on dipolar systems with rotation \cite{diprot1,diprot2}, although no giant vortex was reported.}
 To avoid a collapse of the strongly dipolar atoms, 
we used an improved mean-field
 model, where we include  a Lee-Huang-Yang-type (LHY) interaction \cite{lhy}, appropriate  for dipolar atoms \cite{qf1,qf3,qf2}. 
 As the  dipolar interaction increases beyond a critical value \cite{ex1,ex2},  a strongly dipolar BEC  collapses in the mean-field  Gross-Pitaevskii (GP) model.
However, if 
 a LHY   interaction \cite{qf1,qf2} is  included in the same model, the collapse can be  avoided and the  dipolar BEC can be stabilized 
\cite{expt}.
  The same model  including the LHY interaction was used recently in the study of the formation of droplets in a strongly dipolar BEC  of dysprosium and erbium atoms \cite{expt} { as well as the 
 formation of lattice \cite{luisx,adx} and other structures \cite{pfaux,paul} in that system.} In this study,
 the number of atoms is taken as $N=200000$ and 
 the scattering length $a$ is taken to be  $a=80a_0$, { close to the previously employed    estimates $92(8)a_0$\cite{scatmes}, $87(8)a_0$ \cite{scatmes1} and  $a=88a_0$ \cite{luisx},}  
 where $a_0$ is the Bohr radius.
  The angular frequency $\omega_z$ of the confining  trap in the direction of  polarization  $z$ of the dipolar atoms  is taken to be  $\omega_z =2\pi \times 167$ Hz, as in some experimental \cite{drop2d,drop2d1} and theoretical \cite{luisx,adx} studies on a dipolar BEC  of $^{164}$Dy atoms,   whereas the angular  frequencies
    of the confining  circularly-symmetric trap in  the transverse $x$ and $y$ directions  are taken to be  $\omega_x= \omega_y\equiv \omega_\rho = 0.75 \omega_z$.  The number of atoms, scattering length and the trap parameters for the formation of a giant vortex in the strongly-dipolar BEC of $^{164}$Dy atoms 
 are quite similar to those used in recent  experiments  \cite{expt,drop2d,drop2d1}. 
     The angular velocity of rotation is taken in the range     $0< \Omega  < 0.5\omega_z$ 
   which,   for  $\omega_\rho = 0.75 \omega_z$, is equivalent to $\Omega < 2\omega_\rho/3$, which is not  very high  as it can go as high as $\Omega\to \omega_\rho$ \cite{2c} for  an ordered rotation of the BEC.  For this set of parameters 
the ground state of a strongly-dipolar nonrotating $^{164}$Dy BEC naturally develops  a  hole along the $z$ axis \cite{cyl} as in a Mexican-hat potential \cite{gv0,gv1,gv8}
to have a hollow cylindrical shape  and when set into rotation this BEC   can   accommodate   
 a  stationary giant vortex of angular momentum up to $l=5$.

It is pertinent to ask what is so different in the strongly dipolar $^{164}$Dy BEC, from a nondipolar BEC, considered so far in search of a giant vortex, 
that will  stabilize a giant vortex. The combined action of the long-range non-local dipolar interaction and the collapse inhibiting LHY interaction in the improved mean-field model allows the formation of giant vortices in a harmonically-trapped strongly dipolar BEC. 
The in-plane dipolar
repulsion  in the $x-y$ plane  and the centrifugal repulsion in the rotating condensate create a hollow region along the $z$ direction in a giant vortex, which is  stabilized by the harmonic trap and the strongly repulsive LHY interaction. The harmonic trap  stops the outward expansion of the giant vortex and the strongly repulsive LHY interaction stops its inward collapse, thus stabilizing the giant vortex. 
Although a multiple-charge vortex is unstable in the presence of matter, it
 can be stabilized in this central hollow region as found also in a BEC with an effective  Mexican hat type potential \cite{gv0,gv1,gv8} or with a bubble trap \cite{bbl} responsible to create a central hollow region.

In Sec. \ref{II}  a decription of the the  mean-field model, with the appropriately modified \cite{qf1,qf2}  LHY interaction \cite{lhy},  for a rotating dipolar BEC,  with repulsive atomic interaction (positive scattering length),  is presented. A variational estimate of the energy of the rotating dipolar BEC is also given. In Sec. \ref{III}  we display   numerical results for the formation of stationary giant vortices in strongly dipolar BECs as  obtained by imaginary-time propagation.  For the chosen set of parameters we could stabilize a giant vortex of angular mometum $l \le 5$.  The numerical energy of the giant vortices 
follows a simple theoretical formula \cite{2c}.
The dynamical stability of the giant vortices was established by a steady real-time propagation over a long interval of time of a giant-vortex, after introducing a small perturbation,   using the converged imaginary-time wave function as the initial state. 
 In Sec. \ref{IV} we present a brief summary of our findings.

\section{Improved mean-field model}

  \label{II}

The  improved 
GP equation for a dipolar BEC of $N$ atoms,  of mass $m$ each,
after  the inclusion of  the  LHY interaction, appropriate for dipolar atoms,  can be written as \cite{dipbec,blakie,dip,yuka}
\begin{align}\label{Eq.GP3d}
 \mbox i \hbar \frac{\partial \psi({\bf r},t)}{\partial t} &=\
{\Big [}  -\frac{\hbar^2}{2m}\nabla^2
+V({\bf r})
+ \frac{4\pi \hbar^2}{m}{a} N \vert \psi({\bf r},t) \vert^2 \nonumber\\
& +\frac{3\hbar^2}{m}a_{\mathrm{dd}}  N
\int U_{\mathrm{dd}}({\bf R})
\vert\psi({\mathbf r'},t)\vert^2 d{\mathbf r}'  \nonumber\\
& +\frac{\gamma_{\mathrm{LHY}}\hbar^2}{m}N^{3/2}
|\psi({\mathbf r},t)|^3
\Big] \psi({\bf r},t), 
\\
U_{\mathrm{dd}}({\bf R}) &= \frac{1-3\cos^2 \theta}{|{\bf  R}|^3},\quad a_{\mathrm{dd}}=\frac{\mu_0 \mu^2 m }{ 12\pi \hbar ^2},
\end{align}
where  $V({\bf r}) = \frac{1}{2}m[\omega_\rho^2 (x^2+y^2)+\omega_z ^2z^2]$ is an axially-symmetric trap, 
$\mu$ is the magnetic dipole moment of an atom, 
 $\mu_0$ is the permeability of vacuum, 
 $U_{\mathrm{dd}}({\bf R})$  is the anisotropic dipolar interaction between two atoms placed at $\bf r \equiv \{x,y,z\}$ and $\bf r' \equiv \{x',y',z 
'\}$  
and $\theta$ is the angle between the vector   $\bf R\equiv r-r'$ and   
the $z$ axis, 
$a_{\mathrm{dd}}$ 
is the dipolar length which 
measures  the strength of atomic dipolar interaction;
the wave function  normalization is  $\int \vert \psi({\bf r},t) \vert^2 d{\bf r}=1.$

The coefficient of  the  LHY interaction \cite{lhy}  $\gamma_{\mathrm{LHY}}$ in  Eq. (\ref{Eq.GP3d}), as modified for dipolar atoms \cite{qf1,qf2}, is given by \cite{qf1,qf3,qf2}
$\gamma_{\mathrm{LHY}}= \frac{128}{3}\sqrt{\pi a^5} Q_5(\varepsilon_{\mathrm{dd}}),$ $\varepsilon_{\mathrm{dd}}= \frac{ a_{\mathrm{dd}}}{a},$
where 
$ Q_5(\varepsilon_{\mathrm{dd}})=\ \int_0^1 dx(1-\varepsilon_{\mathrm{dd}}+3x^2\varepsilon_{\mathrm{dd}})^{5/2}$
is  the auxiliary function.
 For nondipolar 
  atoms $ \varepsilon_{\mathrm{dd}} =0,$ while $Q_5(\varepsilon_{\mathrm{dd}} )=1,$ and we get back the usual LHY 
coefficient \cite{lhy}.
The auxiliary function 
{  can be approximated as   \cite{blakie}
\begin{align}\label{exa}
Q_5(\varepsilon_{\mathrm{dd}}) &=\
\frac{(3\varepsilon_{\mathrm{dd}})^{5/2}}{48}   \Re \left[(8+26\eta+33\eta^2)\sqrt{1+\eta}\right.\nonumber
\end{align}
\begin{align}
& + \left.
\ 15\eta^3 \mathrm{ln} \left( \frac{1+\sqrt{1+\eta}}{\sqrt{\eta}}\right)  \right], \quad  \eta = \frac{1-\varepsilon_{\mathrm{dd}}}{3\varepsilon_{\mathrm{dd}}},
\end{align}
subject to the limiting value $Q_5(1)= 3\sqrt 3/2$,
where $\Re$ is the real part. 
In this paper we  will use the analytic approximation (\ref{exa}) for $Q_5(\varepsilon_{\mathrm{dd}})$.


The Hamiltonian of the rotating dipolar BEC in the rotating frame is given by \cite{LF}
$ H=H_0 -\Omega l_z,$
  where $H_0$ is the Hamiltonian in the inertial frame and $l_z\equiv \mathrm{i}\hbar (y\partial /\partial x-x\partial /\partial y)$ is the $z$ component of angular momentum and $\Omega$ the angular velocity of rotation. Using this transformation, the mean-field
GP equation (\ref{Eq.GP3d}) for the trapped BEC in the rotating frame for
$\Omega < \omega_\rho$ can be written as 
\begin{align}\label{eq.GP3d}
 \mbox i \hbar \frac{\partial \psi({\bf r},t)}{\partial t} &=\
{\Big [}  -\frac{\hbar^2}{2m}\nabla^2
+V({\bf r})
+ \frac{4\pi \hbar^2}{m}{a} N \vert \psi({\bf r},t) \vert^2 \nonumber\\
& +\frac{3\hbar^2}{m}a_{\mathrm{dd}}  N
\int U_{\mathrm{dd}}({\bf R})
\vert\psi({\mathbf r'},t)\vert^2 d{\mathbf r}'  \nonumber\\
& +\frac{\gamma_{\mathrm{LHY}}\hbar^2}{m}N^{3/2}
|\psi({\mathbf r},t)|^3-\Omega l_z
\Big] \psi({\bf r},t).
\end{align}
For $\Omega > \omega_\rho$ a harmonically-trapped
rotating BEC makes a quantum phase transition to a non-superfluid state, where a mean-field description of the rotating BEC might not be valid \cite{2c}. 

To write Eq.  (\ref{eq.GP3d})   in the following   dimensionless form  we  express  lengths in units  of the scale    $l_0 = \sqrt{\hbar/m\omega_z}$, time in units of $t_0=\omega_z^{-1}$,  angular frequency in units of $\omega_z$,  energy in units of $\hbar\omega_z$
and density $|\psi|^2$ in units of $l^{-3}$
\begin{align}\label{GP3d2}
\mbox i \frac{\partial \psi({\bf r},t)}{\partial t} & =
{\Big [}  -\frac{1}{2}\nabla^2
+{\frac{1}{2}}\left[{\omega_\rho^2}(x^2+ y^2)+z^2\right]
\nonumber\\ &+ 4\pi{a} N \vert \psi({\bf r},t) \vert^2
+3a_{\mathrm{dd}}  N 
\int 
U_{\mathrm{dd}}({\bf R})
\vert\psi({\mathbf r'},t)\vert^2 d{\mathbf r}'   \nonumber \\ 
&+\gamma_{\mathrm{LHY}}N^{3/2}
|\psi({\mathbf r},t)|^3-\Omega l_z
\Big] \psi({\bf r},t).
\end{align} 
Here and in the following, without any risk of  confusion,  we are representing both the dimensional  and the dimensionless  variables by the same symbols. 
 Equation (\ref{GP3d2})  
can be obtained from   the variational principle 
\begin{align}
\mbox i \frac{\partial \psi}{\partial t} = \frac{\delta E}{\delta \psi^*}, 
\end{align}
from which we obtain the following expression for the energy functional representing the energy per atom 
in a stationary giant vortex in the rotating dipolar BEC
\begin{align}\label{energy}
E &= \int d{\bf r} \left[ \frac{1}{2}{|\nabla\psi({\bf r})|^2} +\frac{1}{2}\left\{{\omega_\rho^2}(x^2+ y^2)+z^2\right\}|\psi({\bf r})|^2  \right. \nonumber  \\
&+ \frac{3}{2}a_{\mathrm{dd}}N|\psi({\bf r})|^2 
\left. \int U_{\mathrm{dd}}({\bf R} )
|\psi({\bf r'})|^2 d {\bf r'} \right. \nonumber \\
& + 2\pi Na |\psi({\bf r})|^4 +\frac{2\gamma_{\mathrm{LHY}}}{5} N^{3/2}
|\psi({\bf r})|^5\nonumber \\
& \left.
-{\mathrm{i}}\Omega \psi^*({\bf r}) \left(y\frac{\partial}{\partial x} -x  \frac{\partial}{\partial y} \right) \psi(\bf r)
\right].
\end{align}

\section{Numerical Results}
\label{III}

 The  partial differential GP equation (\ref{GP3d2}) is solved numerically  employing the split-time-step Crank-Nicolson method
\cite{crank},    using the available   
 C and  FORTRAN  
  programs \cite{dip}. 
and their open-multiprocessing \cite{ompF,omp} versions. 
The imaginary-time propagation  was
used to find the stationary giant vortex state and the real-time propagation to test its dynamical stability.   The
space steps employed for the solution of Eq. (\ref{GP3d2})    are $dx = dy =
0.1$ and $dz = 0.125$,  and the corresponding
time steps in imaginary- and real-time propagations are   $dt = 0.1\times (dx dy dz)^{2/3}$
and $dt = 0.0125\times (dx dy dz)^{2/3}$, respectively.

  For the appearance of  a giant vortex,  we need a strongly dipolar atom with 
  {$a_{\mathrm{dd}}>a$}. For   $a_{\mathrm{dd}}<a$, the system becomes repulsive and behaves more like a nondipolar BEC and no giant vortex can be formed.
  Although,  $a_{\mathrm{dd}}=130.8a_0$ for  $^{164}$Dy atoms, 
we have a certain flexibility in fixing the scattering length $a$, as the scattering length can be modified employing 
the Feshbach resonance technique \cite{fesh} by manipulating an external electromagnetic field.
In this study 
 we take the  scattering length in the domain $86a_0> a> 80a_0$,
 { close to the previously  used values  $a=92(8)a_0$  \cite{expt,scatmes,a01} and $a=88a_0$ \cite{luisx}. }
 For smaller values of $a$ ($\lessapprox 80a_0$)  these states are unstable and collapses, whereas for 
 larger  $a$ ($\gtrapprox 87a_0$) these states decay to multiple states of unit angular momentum.
 With the reduction of contact repulsion, using a Feshbach resonance \cite{fesh}, 
 the choice of scattering length in the region $86a_0> a> 80a_0$  has the advantage of slightly increasing the net attraction, which is found to  facilitate the formation of  a  giant vortex upon rotation. The number of atoms will be taken as $N=200000$,
which favors the formation of  giant vortices.  For   $N \lessapprox  140000$   these states become unstable and can not be found in imaginary-time propagation.
For dysprosium atoms   $m(^{164}$Dy)    $\approx 164 \times 1.66054\times 10^{-27}$ kg.  Consequently, for $\omega_z =2 \pi \times  167$ Hz, used in this study as in some recent experiments  with $^{164}$Dy atoms \cite{drop2d,drop2d1},  the unit of length is $l_0=\sqrt{\hbar/m\omega_z}= 0.6075$ $\mu$m and the unit of time $t_0 = 0.9530$ ms.   It is
true that the angular frequency $\omega_z$ is the same in both cases, but the quasi-2D (two dimensional) trap shape in Refs. \cite{drop2d,drop2d1} is
very different from the trap in the present case ($\omega_\rho = 0.75\omega_z$),  with a much larger  trap frequency  in the transverse direction,
resulting in a different configuration. 
If the angular frequency of the transverse  trap is  very different from this value
($ 0.75\omega_z$) the giant vortices may not be stabilized.
 The demonstration of a  giant vortex  is best  confirmed by a hollow region in the  integrated 2D density $n_{{\mathrm{2D}}}(x,y)$ defined by 
\begin{align}
n_{2D}(x,y)=\int_{-\infty}^{\infty} dz |\psi(x,y,z)|^2.
\end{align}

 To facilitate the numerical simulation of a giant vortex by imaginary-time propagation we use the following  phase-imprinted Gaussian initial state, as   in Refs. \cite{gv11,72},
\begin{align}\label{initial}
\psi({\bf r})= (x+{\mathrm i}y)^l \times \exp\left[-\frac{x^2+y^2}{2\delta_\rho^2} -\frac{z^2}{2\delta_z^2}\right], 
\end{align}
where $\delta_\rho$ and $\delta_z$ are width parameters of the Gaussian state and $l$ its angular momentum. 
 To find a giant vortex with a fixed angular momentum the angular  frequency of rotation is gradually increased in imaginary-time propagation   till the 
giant vortex state can be obtained.
 The angular momentum of the obtained giant vortex is confirmed from a contour plot of the  
phase of the wave function $\psi(x,y,0)$  in the $x-y$ plane.  The phase jump of $2\pi l$ around a closed contour containing 
the $x=y=0$ point confirms a giant vortex of angular momentum $l$, viz. Eq. (\ref{cir}).

\begin{figure}[t!]
\begin{center}

\includegraphics[width=\linewidth]{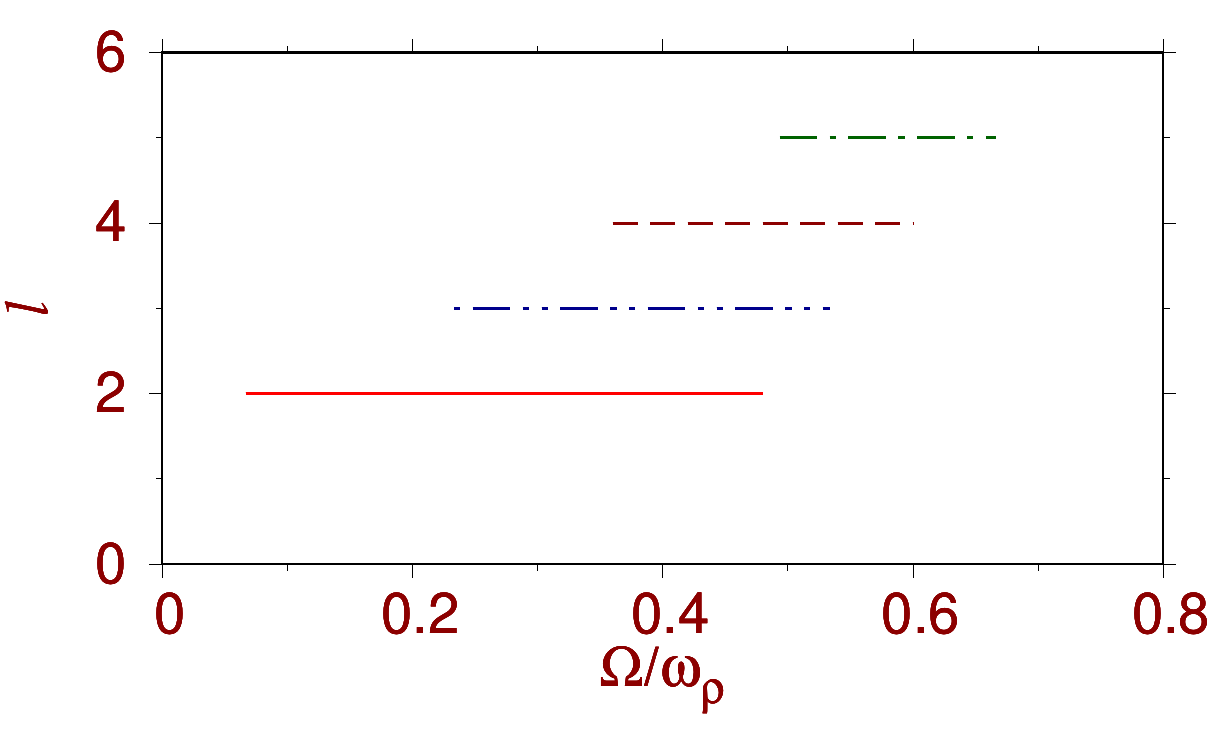}

{
\caption{ The range of angular frequency of rotation $\Omega/\omega_\rho$ for the appearance of a giant vortex of angular momentum $l$ ($5\ge l\ge 2$).   
The parameters of the model   are $N=200000$, $a=80a_0$, $a_{\mathrm{dd}}=130.8a_0$, $\omega_z = 2\pi \times 167$ Hz, $\omega_\rho = 0.75\omega_z$. The plotted results in all figures are dimensionless. 
}
}
\label{fig1} 
\end{center} 
\end{figure}

\begin{figure}[t!]
\begin{center} 

\includegraphics[width=\linewidth]{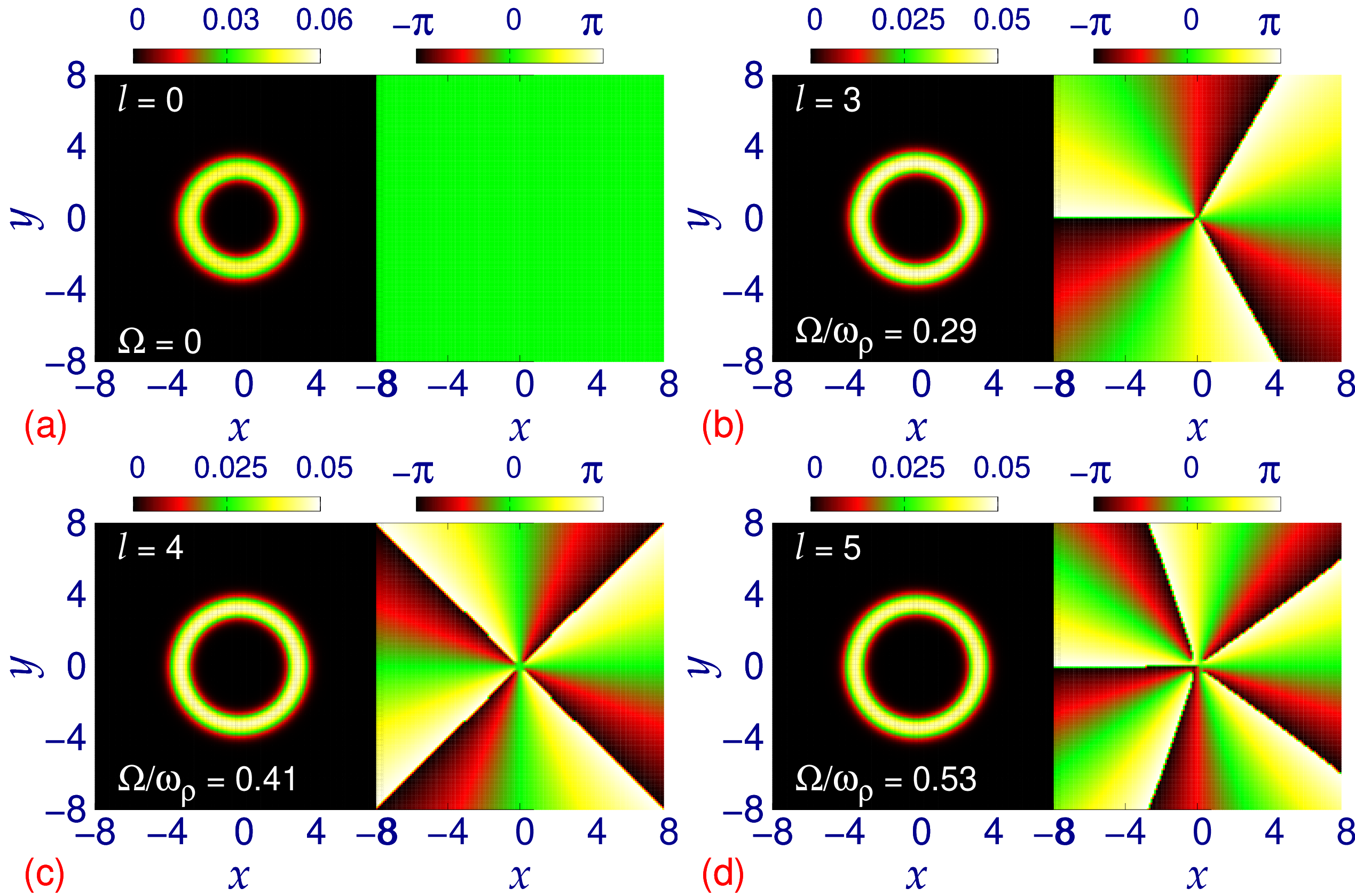}

\caption{Contour plot of dimensionless  2D density 
$n_{{\mathrm{2D}}}(x,y)$  (left panel) and phase of the wave function $\psi(x,y,0)$  (right panel) of the giant-vortices
of angular momentum and angular velocity of rotation (a) $l=0, \Omega=0$,  (b) $l=3, \Omega/\omega_\rho = 0.29$, (c) $l=4, \Omega/\omega_\rho=0.41$,
(d) $l=5, \Omega/\omega_\rho = 0.53$.
The parameters of the model   are $N=200000$, $a=80a_0$, $a_{\mathrm{dd}}=130.8a_0$, $\omega_z = 2\pi \times 167$ Hz, $\omega_\rho = 0.75\omega_z$. 
}
\label{fig2} 
\end{center} 
\end{figure}

 Using the initial function (\ref{initial}) in imaginary-time propagation  we obtain the normal states for angular momentum $l=0,1$ as well as 
 the giant-vortices  of different angular momentum $l=2$ to 5.   As  $l$ is increased, the angular velocity of rotation $\Omega$ is also gradually increased until a converged result is obtained for the giant vortex.  
 Although, the state for $l\le  5$ is a stationary state  with ring topology (a cylindrical state with a coaxial hole) obtainable 
 in a imaginary-time propagation calculation, for larger $l$ this is not so.
{ For larger $l,$ the axially-symmetric hollow cylindrical state breaks into multiple fragments due to large centrifugal force.
A giant vortex with fixed  angular momentum $l$  can be obtained for the rotational frequency $\Omega$ between two limiting values  $\Omega_2  > \Omega > \Omega_1$. This is illustrated in Fig. 1 through a  plot of range of angular frequency of rotation for different  angular molentum $l$ of a giant vortex.  
Each line in this plot shows the extention of $\Omega$ values for which a giant vortex of a specific angular momentum $l$ can be obtained.
 The window of $\Omega $ values $\Delta  \equiv \Omega_2-\Omega_1 $ for stabilizing a giant vortex of angular momentum $l$ reduces as $l$ increases.}
 The smaller  threshold $\Omega_1$  of the rotational frequency for the formation of a giant vortex 
 increases  with the angular momentum $l$.  The $l=1$ vortex   is easily obtained for $\Omega=0.10$ (not shown here).

 The giant vortex states for different $l$  are illustrated through a contour plot of dimensionless 2D density $n_{\mathrm{2D}}(x,y)$ (left panel) and its phase (right panel) in Fig. \ref{fig2}.
All these states including the nonrotating $l=0$ state have a ring topology.
{ Similar ring-shaped  density was obtained in a harmonically-trapped nonrotating dipolar BEC in Ref. \cite{paul} for scattering length $a=100a_0$ ($a/a_{\mathrm{dd}}=0.77$), as well as in Refs. \cite{pfaux,cyl} for  similar sets of parameters.} 
 We illustrate  the results, contour plot of 
 quasi-{2D} density $n_{\mathrm{2D}}(x,y)$ (left panel)  and phase (right panel)  
 of the giant  vortices in Fig.  \ref{fig2}    for  (a) $l=0, \Omega=0,$   (b) $l=3, \Omega/\omega_\rho = 0.29$, (c) $l=4, \Omega/\omega_\rho= 0.41$,
(d) $l=5, \Omega/\omega_\rho = 0.53$. 
The central hollow region 
of the giant vortex  is very prominent in this figure.
We find that the radius of the central hollow region increases monotonically as the angular momentum of the giant vortex  increases.   In these plots, the phase drop of the wave function $\psi(x,y,0)$   along a closed path including the origin  is $2\pi l$ for a giant vortex of angular momentum $l$.

\begin{figure}[t!]
\begin{center} 
\includegraphics[width=\linewidth]{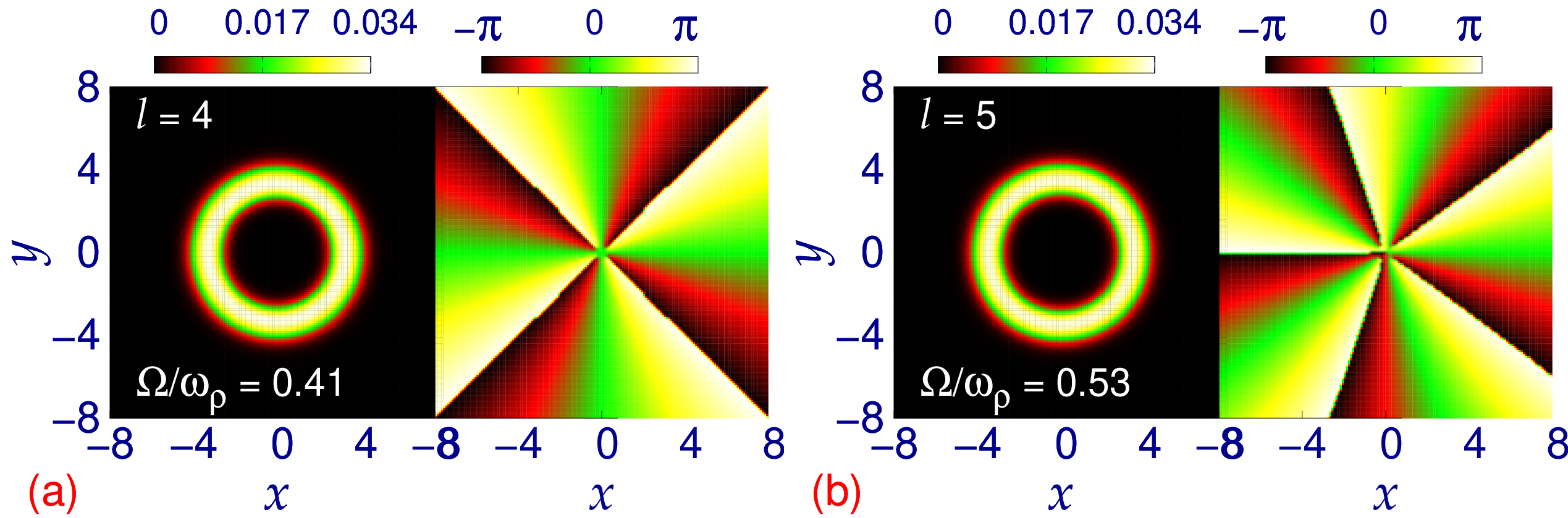}

\caption{Contour plot of dimensionless  2D density $n_{{\mathrm{2D}}}(x,y)$  (left panel) and phase of the wave function $\psi(x,y,0)$  (right panel) 
 of the giant vortices
of angular momentum and angular velocity of rotation (a) $l=4, \Omega/\omega_\rho=0.41$,  (b) $l=5, \Omega/\omega_\rho=0.53$.
The parameters of the model  are $N=200000$, $a=85a_0$, $a_{\mathrm{dd}}=130.8a_0$, $\omega_z = 2\pi \times 167$ Hz, $\omega_\rho = 0.75\omega_z$. 
}
\label{fig3} 
\end{center} 
\end{figure}

 In   Fig. \ref{fig2} the  inner hollow region of the giant vortex  obtained with $N=200000, a=80a_0, \omega_z=2\pi \times 167$ Hz, $\omega_\rho =0.75\omega_z$ is quite sharp and pronounced and  its radius increases monotonically with angular momentum $l$.
 It is of  utmost interest to investigate how the  formation of the giant vortex changes  as the parameters of the model are modified.
As the scattering length $a$ is increased,  the contact repulsion increases,  but   
the inner radius of the giant vortex remains practically the same  as can be found from  the contour plot of the  $l=4,5$  giant vortices  in Fig. \ref{fig3} for  $N=200000, a=85a_0, \omega_z=2\pi \times 167$ Hz,  by comparing with the contour plot  in Fig. \ref{fig2} for $a=80a_0$, with other parameters unchanged.  
Incidentally, the value $a=88a_0$ is a reliable estimate of the scattering length used in many studies 
\cite{luisx,adx}. 
With further increase of $a$, 
for $a\gtrapprox 85a_0$, the giant vortices with large $l$ gradually cease to exist.  For example, for $a=87a_0$,
only the giant vortices with $l\le 4$ are found to exist.   For even larger $a$, for example, for $a=92a_0$,
the contact repulsion increases and, consequently, 
the giant vortex state loses the shell-shaped structure and 
a Gaussian-type BEC in the shape of a normal  solid cylinder   with only $l=1$ vortices 
are  obtained in imaginary-time propagation (not explicitly shown here).

\begin{figure}[t!]
\begin{center} 
 \includegraphics[width=\linewidth]{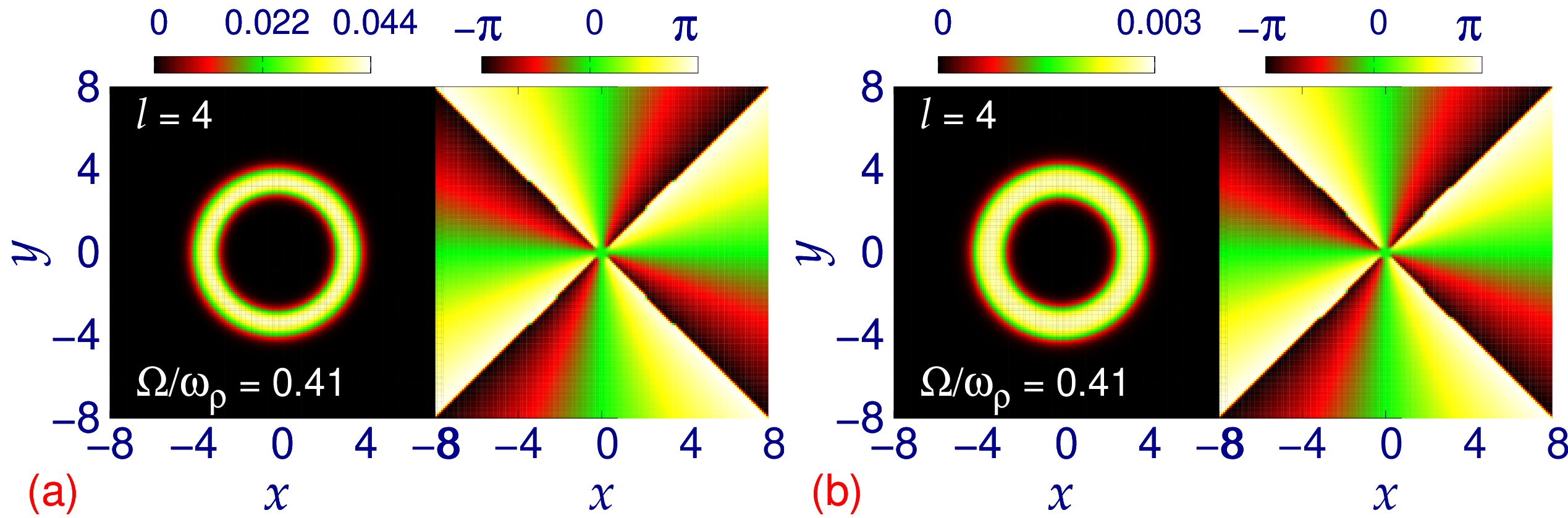}

\caption{Contour plot of dimensionless  2D density 
$n_{{\mathrm{2D}}}(x,y)$  (left panel) and phase of the wave function $\psi(x,y,0)$  (right panel) of the giant vortices of angular momentum $l=4$ for 
 (a)  $N=250000, \omega_z=2\pi \times 167$ Hz  (b) $N=200000, \omega_z=2\pi \times 200$ Hz.
Other parameters of the model  are the  same as in Fig. \ref{fig2}.
}
\label{fig4} 
\end{center} 
\end{figure}

We next investigate the fate of the giant vortices as other parameters of the model are changed. If $N$ is decreased below a critical value, the cylindrical-shell-shaped states are not possible and only fragmented droplet-type states appear. For $\omega_z=2\pi \times 167$ Hz this happens for  $N\approx 100000$ (not explicitly shown here).  However, the axially-symmetric giant vortices survive for larger $N$ as illustrated in Fig. \ref{fig4}(a) for $N=250000$ with all other parameters unchanged.  Finally, we study the fate of the giant vortices as the axial frequency $\omega_z$ is changed.  As $\omega_z$ is reduced further, these states disappear, for example for $\omega_z=2\pi \times 100$ Hz, when multi-fragment states appear (not shown here). However, these states survive for larger values of $\omega _z$  as illustrated in Fig. \ref{fig4}(b) for  $\omega_z=2\pi \times 200$ Hz with other parameters unchanged.  However, if the transverse trapping frequency $\omega_\rho$   is reduced below $\omega_\rho =0.60\omega_z,$ or increased above 
$\omega_\rho =0.90\omega_z $,  these states disappear. In the former case, the system is no longer strongly dipolar  and in the latter case the stronger   trapping frequency transforms the hollow cylindrical states to the shape  of a solid cylinder.

\begin{figure}[t!]
\begin{center}

\includegraphics[width=\linewidth]{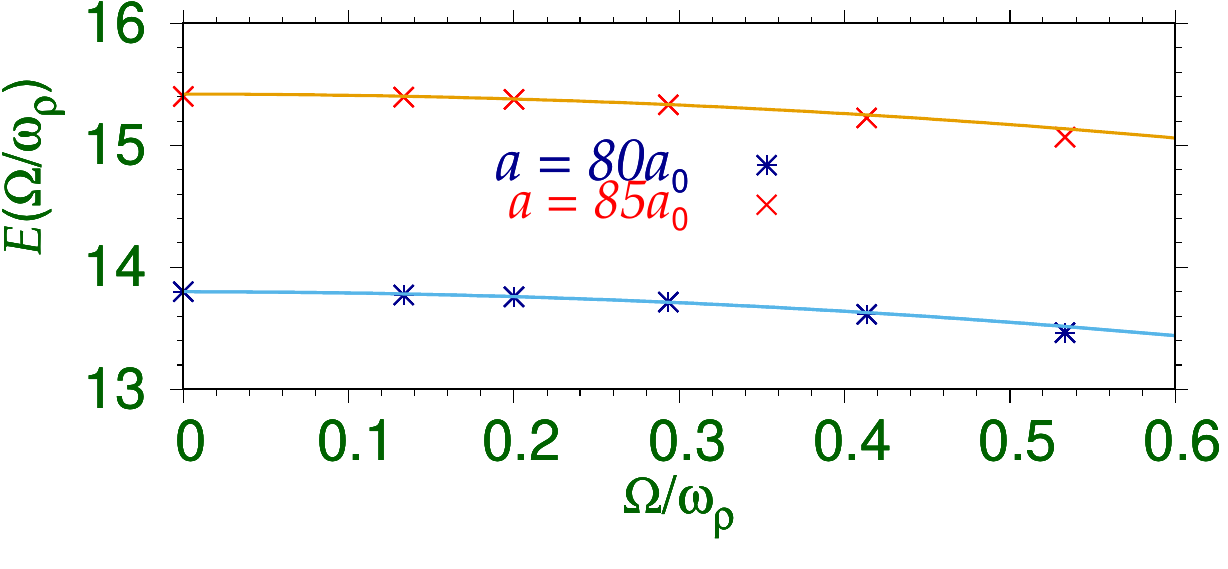}

\caption{  The energy per atom  of the giant vortex of different angular momentum  as a function of the angular velocity of rotation for $a=80a_0$ and $a=85a_0$. The points correspond to different angular-momentum states  ($l=0$ to 5) as obtained from Eq. (\ref{energy}); the increase in angular momentum of the giant vortex correspond to a decrease in energy.  The lines are the   theoretical estimate (\ref{2c}) with $I= 2$.}
\label{fig5} 
\end{center} 
\end{figure}

Fetter made the following  theoretical estimate of the energy per atom 
of a nondipolar dilute rotating BEC  in a harmonic trap \cite{2c} 
\begin{align}\label{2c}
E\left({\Omega}/{\omega_\rho}\right) = E(0) -\frac{I}{2} \left(\frac{\Omega}{\omega_\rho}\right)^2
\end{align}
{where $I$ is the moment of inertia of rigid-body rotation of an atom of the superfluid} and $\frac{1}{2}I (\Omega/\omega_\rho)^2$ is the rotational kinetic energy.
We  have applied this estimate to the present case of giant vortex in a strongly dipolar BEC of $^{164}$Dy atoms including the LHY interaction.
A plot of energy $E(\Omega/\omega_\rho)$  of  Eq. (\ref{energy}) versus the angular velocity of rotation $\Omega/\omega_\rho$ for 
the giant vortices  of  different angular momentum, $l=0$ to 5, is displayed in Fig. \ref{fig5}, for  $a=80a_0$ and $85a_0$, which 
  confirms the universal behavior of the rotational energy given by Eq. (\ref{2c}), {where for $E(0)$ we have taken the numerically obtained energy (\ref{energy}) for $\Omega=0$. The points correspond to the  the different  vortices ($l\le 5$) illustrated in Fig. \ref{fig2};} the  energy decreases with increasing angular momentum $l$ in a monotonic fashion.
 The full lines are  the theoretical estimates (\ref{2c}) for  $I=2$.  { The moment of inertia $I$ was obtained
 from Eq. (\ref{2c}) using the numerical energy for  $\Omega=0$ and another energy for $\Omega \ne 0$.  }
 It is pleasing to see that the Fetter's universal estimate of energy for a dilute BEC works very well for a strongly dipolar dense BEC for two values of the scattering lengths  using the same moment of inertia $I=2$.

\begin{figure}[t!]
\begin{center} 

\includegraphics[width=\linewidth]{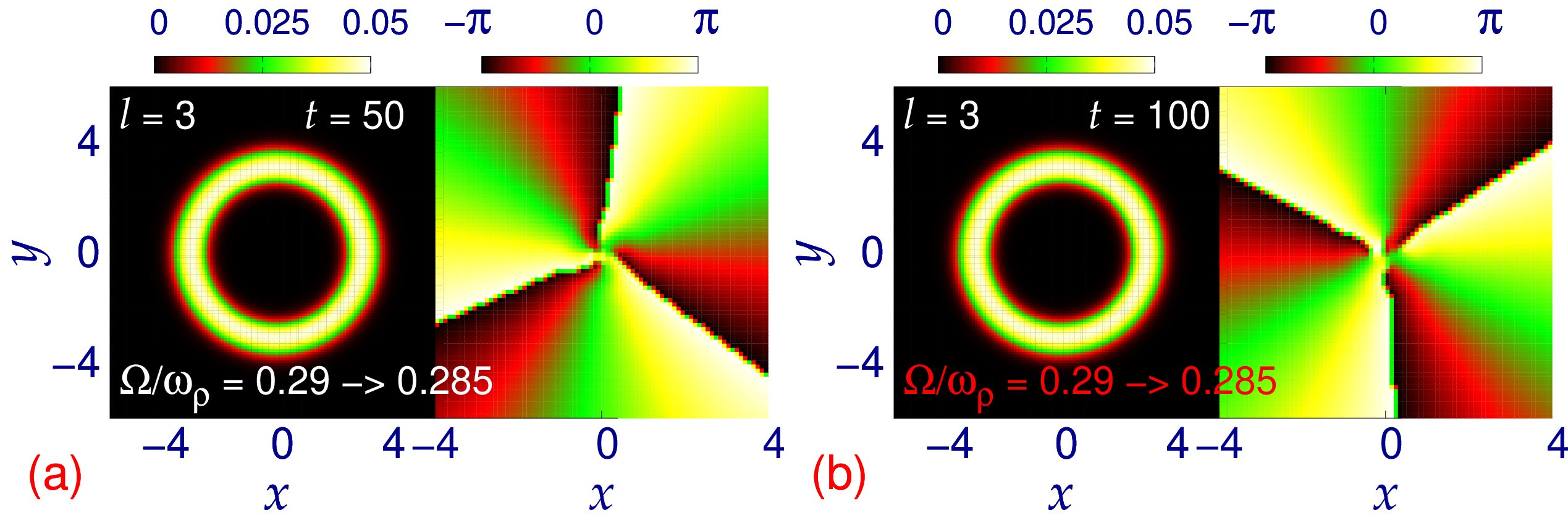}
\includegraphics[width=.7\linewidth]{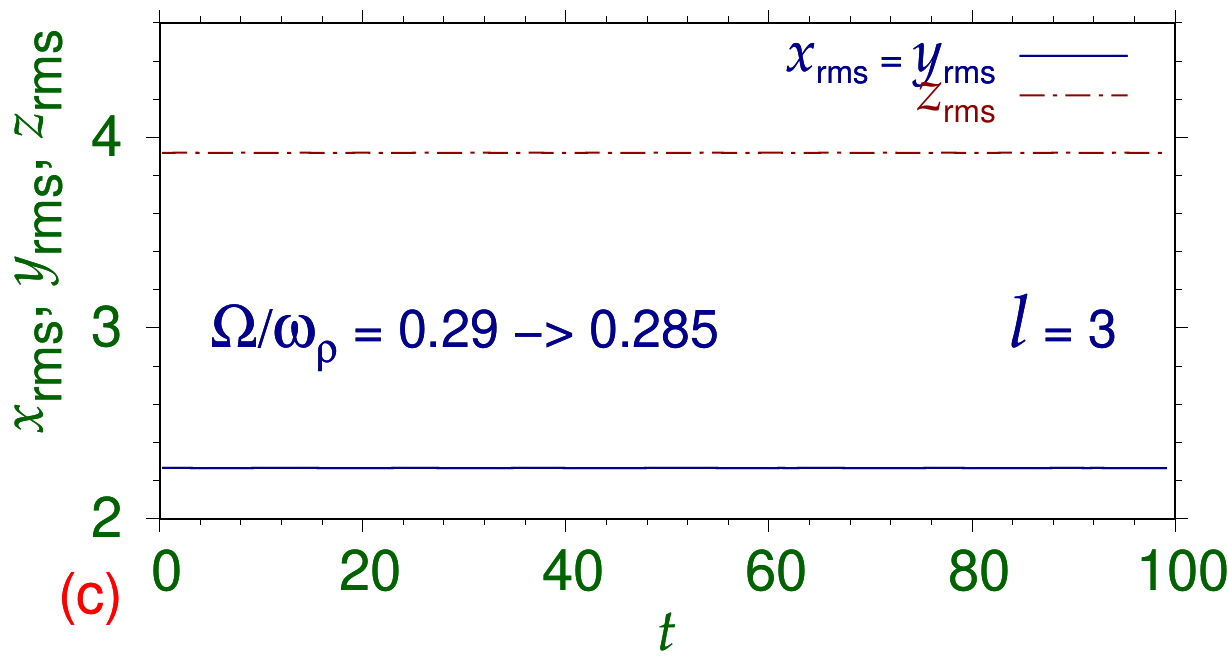}

\caption{The same as in Fig. \ref{fig2}  of the giant vortex 
of Fig. \ref{fig2}(d)  of angular momentum $l=3$ at times  
 (a)  $t=50$  (b) $t=100$ during real-time propagation using the imaginary-time wave function as the inital state after changing the rotational velocity $\Omega/\omega_\rho$ from 0.29 to 0.285. 
 The evolution of the rms sizes $x_{\mathrm{rms}}$, $y_{\mathrm{rms}}$,   and $z_{\mathrm{rms}}$
 versus time for  the dynamics of (c) Figs. \ref{fig6}(a)-(b). 
Other parameters of the model  are the  same as in Fig. \ref{fig2}.
}
\label{fig6} 
\end{center} 
\end{figure} 

 To demonstrate the dynamical stability of the giant vortices,  we consider the giant vortex  of Fig \ref{fig2}(d) with $l=3$, $a=80a_0$ and $\Omega/\omega_\rho =0.29$   and change $\Omega/\omega_\rho$   to $0.285$ in the model equation (\ref{GP3d2}) maintaining other parameters unchanged. Using the converged  imaginary-time wave function  as the initial state, { we perform real-time propagation of Eq.  (\ref{GP3d2}) with  $\Omega/\omega_\rho=0.285$ for 100 units of time (about 100 ms in the present case).} The density and the phase of the real-time wave function at times $t= 50 $ and $100$  are exhibited in Figs. \ref{fig6}(a) and (b).  The phase drop around a closed circular path at the center in (a)-(b)
 is $6\pi$,  corresponding to an angular momentum of $l=3$.  {  The  evolution of the root-mean-square (rms) sizes  during the real-time propagation exhibited in  Figs. \ref{fig6}(a)-(b)   is shown in Figs. \ref{fig6}(c).
  The circularly-symmetric density distribution and the correct phase drop at the center in Figs. \ref{fig6}(a) and (b) at $t=50$ and  $t=100$ as well as the steady oscillation of the rms sizes  with a very small amplitude in Figs. \ref{fig6}(c)
   guarantee the dynamical stability of the giant vortices.  As we do not know the different loss rates of the dipolar BEC,  we did not include any dissipation in this study. A demonstration of the dynamical stability including appropriate dissipations is beyond the scope of the present investigation, and would be the subject of a future investigation.}

{
    It has been
demonstrated that ring-shaped densities with nondipolar \cite{persis-nd} and dipolar \cite{persis-dip} condensates  can support multiply
charged circulation even at zero rotation frequency, called persistent currents,
 that differ from vortices as their existence do not require rotation,
contrary to vortices.  Typically, a persistent current was generated by imprinting a  vortex on a stable ring-shaped condensate \cite{persis-nd} in an appropriate (non-harmonic) trap. We verified that such persistent current is not possible in the present condensate  in the absence of rotation. To demonstrate we considered the non-rotating condensate of Fig. \ref{fig2}(a) in a harmonic trap 
and imprinted its wave function with a multiply-charged vortex and performed real-time simulation 
with this initial state for $\Omega=0$. 
The multiply-charged vortex is immediately destroyed demonstrating the absence of persistent current in the present case. Also, we have seen that in imaginary-time propagation the giant vortices do not exist for $\Omega=0$, viz. Fig.  1,   corroborating the fact that the present $l>1$ states are giant vortices.}
 
\section{Summary} 
\label{IV}
  
In conclusion, we have demonstrated the possibility of generating a dynamically stable giant vortex of angular momentum $l=2$ to 5 in a 
  harmonically-trapped strongly-dipolar BEC of $^{164}$Dy atoms, for parameters similar to those used in recent studies \cite{expt,drop2d,drop2d1} on this system, by employing imaginary- and real-time propagation of   an improved mean-field equation including a LHY interaction \cite{lhy} appropriately modified \cite{qf1} for dipolar atoms. The giant vortices appear for number of atoms $N$ in the range $140000 \lessapprox  N \lessapprox 250000$,  angular trap frequency in the range $2\pi \times 130$ Hz $\lessapprox \omega_z \lessapprox 2\pi \times 230$ Hz 
and   $0.60 \lessapprox  \omega_\rho/\omega_z \lessapprox  0.90$ and for scattering length in the range
$80a_0 \lessapprox a \lessapprox 85a_0$, viz. Figs. \ref{fig3} and \ref{fig4}.
A large dipole moment is essential for the formation of  a strongly-dipolar BEC necessary for generating a giant vortex.    The energy of the giant vortices (\ref{energy}) are in agreement with the theoretical estimate (\ref{2c}) \cite{2c}, viz. Fig. \ref{fig5}. 
The dynamical stability of the giant vortices was established by real-time propagation upon a small perturbation of the   trap frequency,  
employing the converged imaginary-time wave function as the initial state.
 The present study may initiate new avenues of theoretical and experimental research in the area of cold atoms. 
\section*{CRediT authorship contribution statement}
L. E. Young-S. and S. K. Adhikari: Conceptualization, Methodology, Validation, Investigation, Writing – original draft, Writing – review and  editing,
Visualization.

\section*{Declaration of competing interest}
The authors declare that they have no known competing financial interests or personal relationships that could have appeared
to influence the work reported in this paper.
\section*{Data availability}
No data was used for the research described in the article.

\section*{Acknowledgments}

 LEY-S. acknowledges partial  financial support by the Vicerrectoría de Investigaciones - Universidad de Cartagena, Colombia through Projects  064-2023. 
SKA acknowledges partial support by the CNPq (Brazil) grant 301324/2019-0. 
 The use of the supercomputing  cluster of the Universidad de Cartagena is acknowledged.




\end{document}